\documentclass[paper]{ite}
\usepackage{graphicx}
\usepackage[usenames]{color}
\usepackage{amssymb}
\usepackage{url}
\usepackage{etoolbox}

\makeatletter
\newcommand{\citerangesep}{\ensuremath{\sim}}
\patchcmd{\@citex}{\~{}}{\noexpand\citerangesep}{}{\PackageWarning{hamamoto_mta}{cite patch 1 failed}}
\patchcmd{\@citex}{\~{}}{\noexpand\citerangesep}{}{\PackageWarning{hamamoto_mta}{cite patch 2 failed}}
\makeatother

\makeatletter
\def\ite@in@ext{pdf}
\def\arxiv@foot{\foot@font Preprint submitted to ITE Transactions on Media Technology and Applications}
\def\ps@ite{\let\@mkboth\@gobbletwo
 \def\@oddhead{\@Ltop\hfil\@Rtop}\let\@evenhead\@oddhead
 \def\@oddfoot{\@Lbot\hskip3.25mm\arxiv@foot\hfil{\nombre@font\thepage}\hskip3.25mm\@Rbot}%
 \let\@evenfoot\@oddfoot}
\let\ps@ITE\ps@ite
\makeatother

\title{Projected Route Cues for Bouldering:\\ Effects of Circular, Linear, and Combined Highlighting on Route Observation and Climbing}

\authorlist{%
 \authorentry{Yuki Uranishi}{n}{osakad3}
 \authorentry{Tamon Hamamoto}{n}{osakad3}
 \authorentry{Chang Liu}{n}{kyotomed}
 \authorentry{Goshiro Yamamoto}{n}{kyoto}
 \authorentry{Masato Kobayashi}{n}{osakad3}
 \authorentry{Photchara Ratsamee}{n}{kansai}
 \authorentry{Haruo Takemura}{n}{osakad3}
}
\affiliate[osakad3]
 {D3 Center, The University of Osaka}
 {Osaka, Japan}
\affiliate[kyotomed]
 {Graduate School of Medicine, Kyoto University}
 {Kyoto, Japan}
\affiliate[kyoto]
 {Kyoto University Hospital}
 {Kyoto, Japan}
\affiliate[kansai]
 {Faculty of Engineering Science, Kansai University}
 {Osaka, Japan}

\videoFlag{0} %

\begin{document}
\begin{abstract}
Before attempting a bouldering problem, climbers inspect the route to find its holds and plan their moves, which novices find difficult on walls where many routes overlap. We built a projector-camera system that projects route cues onto the wall and compared three cues: circles on the route holds, limb-colored lines connecting the holds, and both combined. Ninety-six participants, 61 of them first-time climbers, each climbed four routes, one per cue and one without projection, in a Latin-square order. In mixed-effects models, all three cues shortened the time before the first attempt; with circles, its geometric mean fell from 83\,s to 39\,s. Circles also increased the number of moves made within each trial. Lines led to more use of off-route holds than circles, mainly on routes where such holds lay under the projected lines. Participants preferred the combined cue, although no behavioral advantage of it over circles was detected.

\end{abstract}
\maketitle

\section{Introduction}
\label{sec:introduction}
Bouldering is a form of sport climbing in which climbers ascend an artificial wall approximately 4\,m high using holds placed by a route setter. Among all holds on the wall, climbers may use only a designated subset, referred to as a \textit{bouldering problem}, to reach the goal. Unlike other forms of sport climbing, bouldering is done without ropes on walls or boulders low enough for the climber to jump or fall to the ground, and falls are part of the activity \cite{Gill1969}. Because a fall brings the climber straight to the ground, climbers must optimize their movements based on the shape and arrangement of each hold for both performance and safety.

Before climbing, climbers usually inspect the route from the ground to locate its holds and to plan the sequence of moves. This step is called \textit{observation} in bouldering and route preview in the climbing literature. Sanchez et al.\ \cite{sanchez2012efficacy} studied 29 intermediate to expert climbers on indoor sport routes and found that a 3-min preview did not make climbers more likely to finish a route, but that climbers made fewer and shorter stops during the ascent after a preview. Seifert et al.\ \cite{seifert2017role} recorded the gaze of inexperienced and experienced climbers during preview and found that the number of times a climber scanned the route was correlated with how the climber explored and changed holds during the subsequent ascent. These studies show that observation affects how a route is climbed, even when it does not affect whether the route is completed.

Observation is difficult for novices, who do not yet know which hold shapes are usable or which movements a given arrangement of holds requires. In commercial gyms, many problems are set on the same wall, and the holds of each problem are typically marked by tape labels or by hold color. A novice must therefore first find the holds of the target problem among those of other problems before planning any movement, and may lose track of them again during the climb.

Projecting information onto the wall can indicate which holds belong to a problem without modifying the wall or attaching devices to the climber. Such projection onto physical surfaces is known as spatial augmented reality \cite{Bimber2005}. Previous systems have projected route highlights and replayed recorded climbing movements on climbing walls \cite{augmentedWall,betacube}. These systems showed that route information can be projected onto a climbing wall. However, the cues they project differ in what they convey: a mark on each hold indicates only where the holds are, whereas a line between holds also indicates the order of the holds and the limb to be used. To our knowledge, no controlled study has compared how such differences in cue design affect the time climbers need to observe a route and how they then climb it.

In this paper, to help climbers find the target problem among many and move on to attempting it, we compare three projected cues: circles projected onto the route holds (circular highlighting), lines that connect the route holds in a different color for each limb (linear highlighting), and both together (combined highlighting). We built a portable projector-camera system that maps the cues onto the wall using a depth sensor, and conducted a user study with 96 participants, 61 of whom were bouldering for the first time. Each participant climbed four routes, each under one of the three cues or without projection. From video recordings, we measured the time before the first attempt (which includes observing the route), the number of attempts within the 3-min trial, the number of moves, route completion, and the use of off-route holds. We analyzed these measures with mixed-effects models that account for the four trials of each participant.

The contributions of this study are as follows:
\begin{itemize}
    \item We show that all three projected cues shortened the time before the first attempt. With circular highlighting, the geometric mean of this time fell from 83\,s to 39\,s, and participants made more attempts and more moves within each trial.
    \item We show that linear highlighting led to more use of off-route holds than circular highlighting, and that with linear highlighting these errors occurred mainly on the two routes where off-route holds lay under the lines.
    \item We report that participants preferred the combined cue to circular highlighting, although the combined cue led to a longer time before the first attempt and no behavioral advantage of it over circular highlighting was detected. From these results, we derive design recommendations for projected route cues of these kinds.
\end{itemize}

The remainder of this paper is organized as follows. Section~\ref{sec:related_work} reviews prior work on climbing support. Section~\ref{sec:proposed_method} describes the system and the cues. Section~\ref{sec:user_study} presents the user study, its results, and a discussion. Section~\ref{sec:conclusion} concludes the paper.

\section{Related Work}
\label{sec:related_work}
This section reviews three groups of studies: climbing support that uses sensors or wearable devices, systems that present information on or around a climbing wall by projection or mixed reality, and projected guidance for body movement outside climbing.

\subsection{Sensor-based and Wearable Support for Climbing}
Daiber et al.\ developed a mobile augmented reality application with which climbers create and share boulder problems and challenges with their friends \cite{bouldAR}. Ivanova et al.\ proposed recommending routes to climbers on the basis of their activity recorded automatically by sensors in a gym and of diary data entered in a mobile application \cite{ivanova2021climber}. Liljedahl et al.\ developed Digiwall, a climbing wall whose semitransparent holds contain capacitive sensors that react to hands and feet and LEDs that light the holds; games and challenges on the wall are presented through the lit holds and through sound and music from a surround sound system \cite{digiwall}.
Force sensors built into holds or into the bolts that fix them have been used to measure the forces that climbers apply to holds \cite{Fuss2006fs,Lechner2013fs,Iguma2020fs,hayashida2024bolt}.

Wearable sensors have been used to assess climbing skill \cite{ladha2013climbax}, to recognize which route was climbed \cite{kosmalla2015climbsense}, and to monitor muscle fatigue \cite{dindorf2023vivo}, and the design of wearable systems for sports has been reviewed by Mencarini et al.\ \cite{mencarini2019designing}. Mencarini et al.\ proposed four design considerations for wearable devices that support beginners in learning to climb by augmenting the communication between climbing partners and with the instructor \cite{Mencarini2016}, and deployed a vibrotactile wearable device for communication between instructor and trainees during an indoor climbing lesson \cite{mencarini2019co}. Feeken et al.\ attached a device to climbing shoes that vibrates when foot pressure stays low or the body position changes abruptly \cite{climbAssist}.
Kosmalla et al.\ compared visual, tactile, and auditory notifications during climbing and found sound to be the most suitable channel and light to be inappropriate \cite{climbAware}.

These approaches require sensors on the wall or devices on the climber. Instrumenting holds is difficult in existing gyms, where routes are reset regularly, and a device worn by a climber can be damaged or cause injury when the climber falls.

\subsection{Projection and Mixed Reality on Climbing Walls}
\label{ref:ProjectionMapping}
Extended reality technologies have been applied in many sports \cite{lenoury2022narrative}. For climbing, Wiehr et al.\ developed betaCube, a mobile projector-camera unit that calibrates itself to a bouldering wall, highlights holds selected on a smartphone image, and replays recorded climbing movements at full scale on the wall \cite{betacube}.
Kajastila and H\"am\"al\"ainen projected onto a climbing wall route highlights, routes generated from the holds a climber touched, a life-size image of the climber, and an entertaining animation that followed the climber, and asked climbers in a Wizard-of-Oz study which functions they found useful \cite{augmentedWall}. Kajastila et al.\ later installed an augmented climbing wall that combines computer vision and interactive projected graphics in a commercial climbing center, and evaluated its movement-based games in three user studies \cite{kajastila2016augmented}.
Kosmalla et al.\ projected a life-size video of an instructor onto the wall and compared it with a third-person view projected on the wall and a third-person view on Google Glass; participants preferred the life-size projection \cite{climbVis}.
Overdevest et al.\ presented three-dimensional virtual plants on a bouldering wall through an AR headset and signaled contact with vibrotactile actuators on the forearm \cite{overdevest2025boulderplay}. Climbing has also been studied in mixed reality \cite{tiator2018venga} and in virtual reality environments in which climbers wearing a head-mounted display climb a physical wall \cite{kosmalla2017exploring,schulz2019role,kosmalla2020importance,kosmalla2022infinitywall}.

These studies established that route information can be projected onto a climbing wall without instrumenting the wall or the climber. Their evaluations, however, concerned which functions climbers found useful and how climbers used and enjoyed the systems; they did not compare different ways of drawing the same route, and they did not measure how long climbers observed a route before climbing it.

\subsection{Projected Guidance for Body Movement}
Outside climbing, projection has been used to guide body movement in place. Sodhi et al.\ projected cues onto the user's hand to guide hand movements in three dimensions; participants followed the projected cues with a mean error of 21.6\,mm, about 85\% more accurately than with video instructions \cite{lightGuide}. A ping-pong table has been augmented with graphics and sound that respond to where and how the ball is hit \cite{pingpongplus}, and golf swing training has been supported by projecting a virtual shadow of the learner, overlaid with the outline of an expert, onto the ground \cite{VirtualShadow}. In these systems, the projected information indicates the target of a movement on the surface where the movement takes place. Our study examines the same principle on a climbing wall, where each movement targets a hold and many off-route holds lie close to the route holds.

\section{Projector-Camera System and Projected Cues}
\label{sec:proposed_method}
The contribution of this study is the comparison of projected cue designs and the measurement of their effects on climbing; the projector-camera system was built from existing tools so that the cues could be placed accurately on the holds. Section~\ref{ref:system} describes the system and its calibration, and Section~\ref{ref:highlight} describes the three cues and the information each of them conveys.

\subsection{Projector-Camera System}
\label{ref:system}
\begin{figure}[tb]
    \centering
    \includegraphics[width=0.9\linewidth]{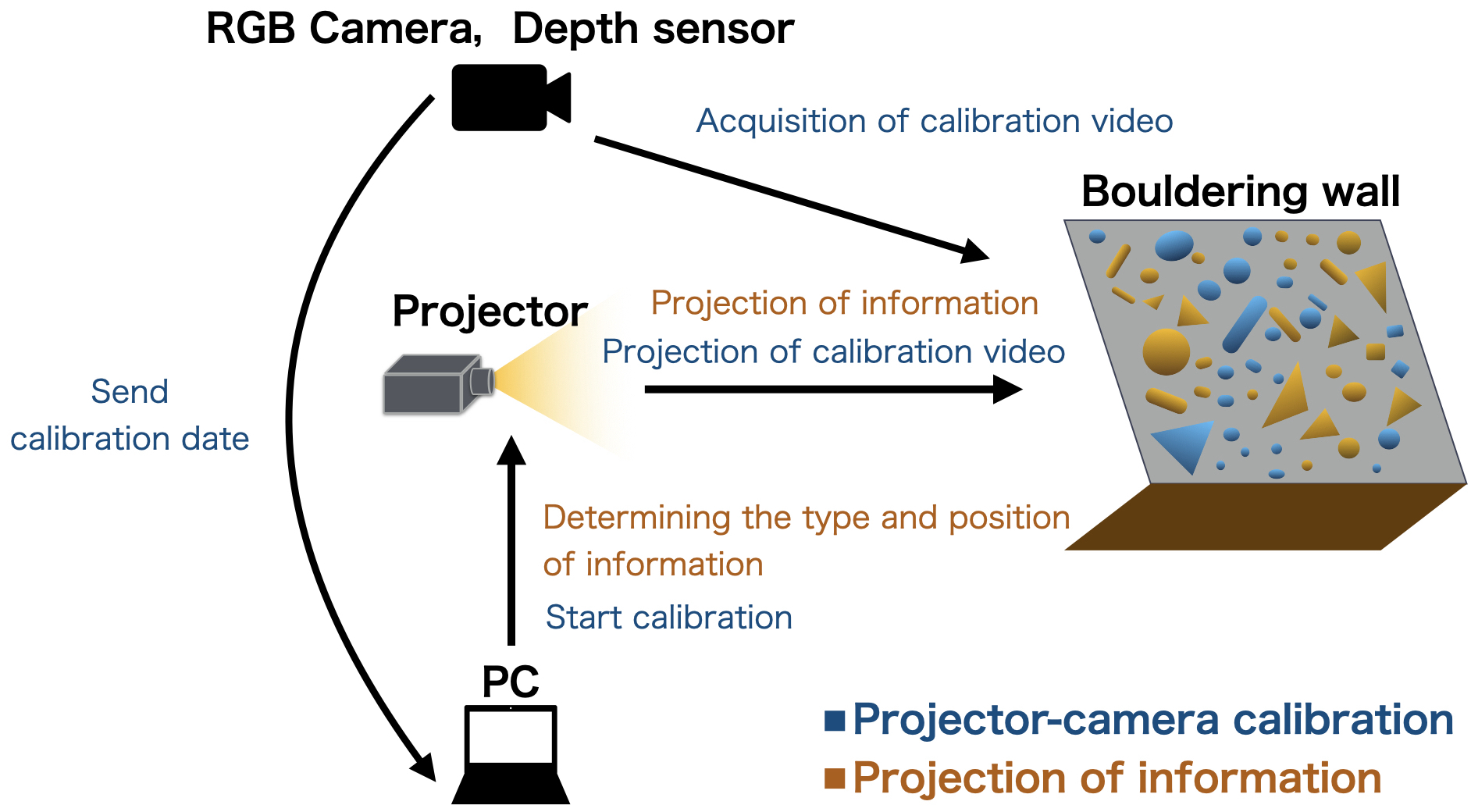}
    \caption{Hardware configuration of the projector-camera system: an RGB camera with a depth sensor and a projector that projects the cues onto the climbing wall.}
    \label{fig:systemConcept}
\end{figure}

Figure~\ref{fig:systemConcept} shows the hardware configuration. A BenQ MW632ST short-throw projector and a Microsoft Kinect V2, which contains an RGB camera and a depth sensor, are mounted on two stacked open racks (each 40\,cm wide, 20\,cm high, 30\,cm deep), with the projector at the bottom and the Kinect at the top. A 3D-printed part raises the rear of the projector by 56\,mm to tilt the projection upward, and the racks are fixed to a wheeled stand for moving the system between walls.

\begin{figure}[!t]
    \centering
    \includegraphics[width=0.9\linewidth]{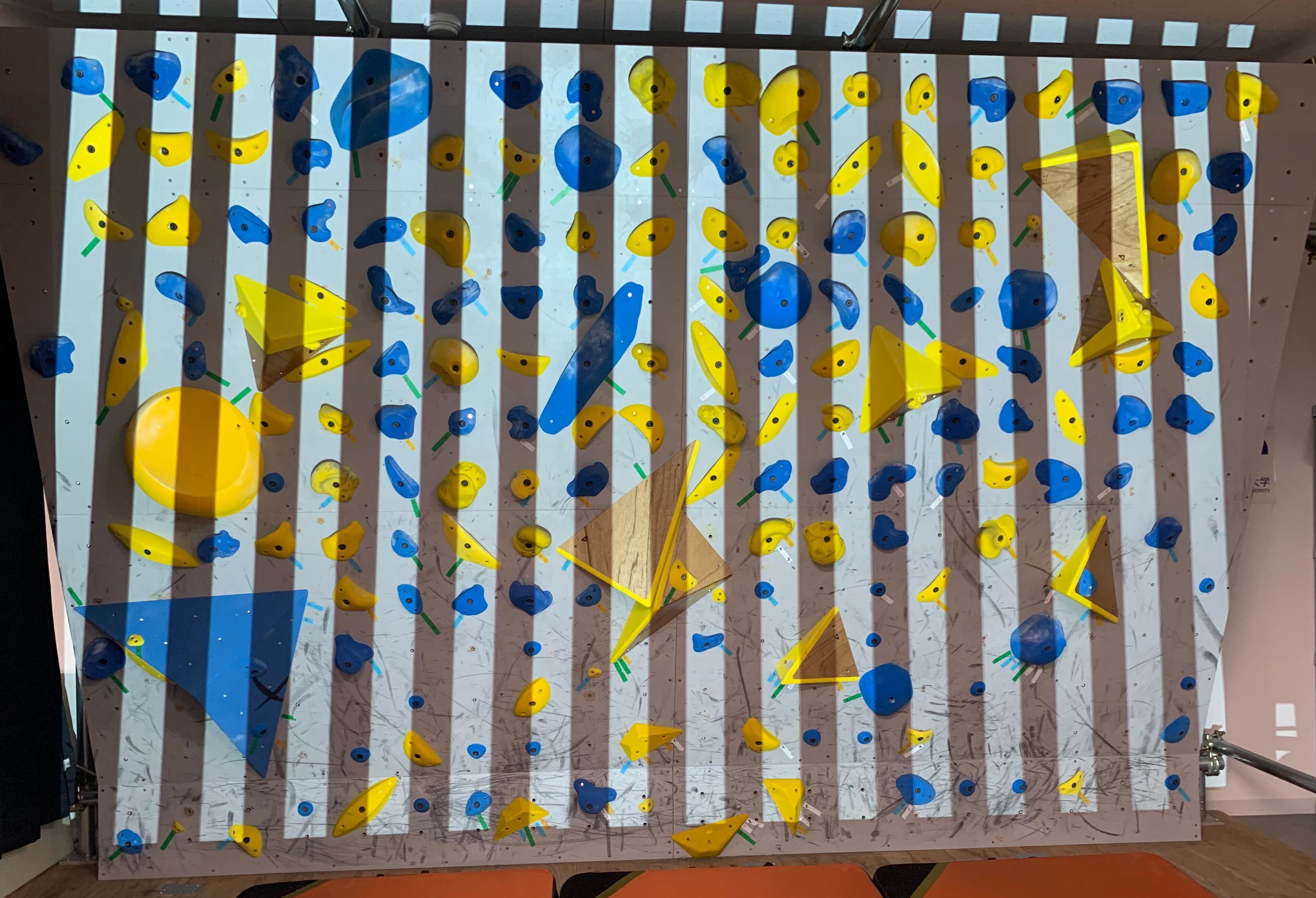}
    \caption{Gray code pattern projected onto the wall for projector-camera calibration.}
    \label{fig:graycode}
\end{figure}

The projector and the camera were calibrated with the ProCamCalibration tool of the RoomAlive Toolkit \cite{RoomAlive}. Horizontal and vertical Gray code patterns (Fig.~\ref{fig:graycode}) are projected onto the wall and captured by the camera, which establishes a pixel-to-pixel correspondence between the projector and the camera \cite{RoomAliveToolkitPaper}. Each corresponding point is assigned a three-dimensional position from the depth image, and the intrinsic and extrinsic parameters of the projector and camera are estimated from these points with OpenCV.

To evaluate the calibration, we projected $\times$ marks onto bolt holes detected by the depth sensor and measured the offset between each mark and the bolt hole with a ruler. In two sessions, 476 marks were measured, 191 on the flat wall and 285 on holds. The reprojection error was $5.64 \pm 2.95$\,mm (mean $\pm$ standard deviation), $5.82 \pm 3.04$\,mm for marks on the flat wall and $5.46 \pm 2.86$\,mm for marks on holds, and the largest error was 16\,mm (Table~\ref{tab:ReprojectionError}). A circle of 155\,mm diameter around a hold of 105\,mm diameter leaves a margin of 25\,mm on each side, larger than every measured error.

\begin{table}[!b]
\centering
    \caption{Reprojection error of the projector-camera calibration.}
    \label{tab:ReprojectionError}
    \scalebox{0.92}{
    \begin{tabular}{ccccc}
    & Marks & Mean [mm] & SD [mm] & Max [mm] \\
    \hline
    All & 476 & 5.64 & 2.95 & 16 \\
    On the wall & 191 & 5.82 & 3.04 & 15 \\
    On holds & 285 & 5.46 & 2.86 & 16 \\
    \hline
    \end{tabular}
    }
\end{table}

\begin{figure}[!t]
    \centering
    \includegraphics[width=0.9\linewidth]{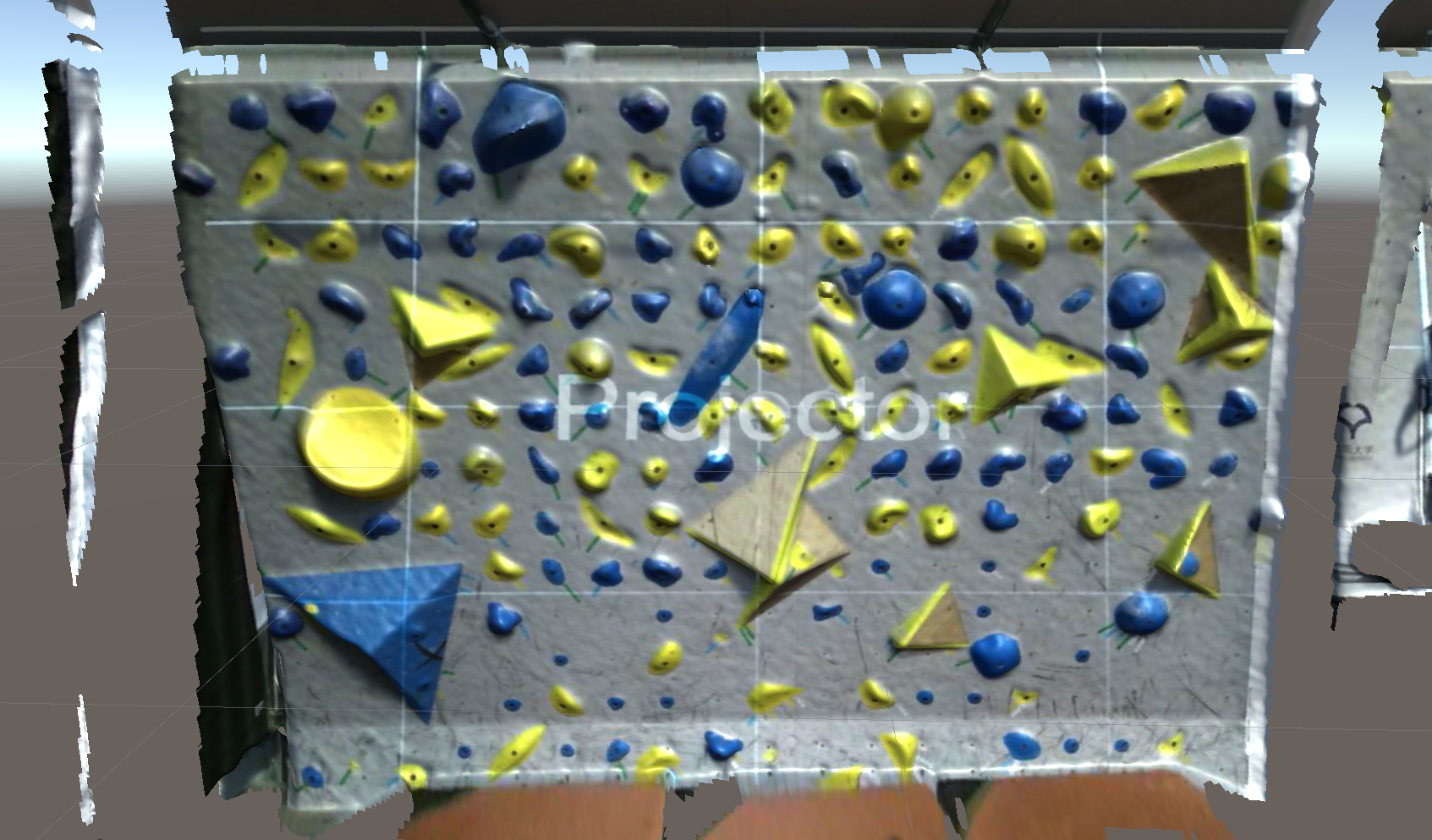}
    \caption{Reconstructed three-dimensional model of the wall in Unity. Cues are placed on this model and projected onto the physical wall.}
    \label{fig:3Dwall}
\end{figure}

The cues are rendered with the RoomAlive Toolkit for Unity. The toolkit reconstructs the three-dimensional shape of the wall, including the holds, from the depth sensor, and the reconstructed model is placed in the Unity scene using the calibration result (Fig.~\ref{fig:3Dwall}). A virtual object placed on a hold in this model is projected onto the corresponding hold on the physical wall.

\subsection{Projected Cues}
\label{ref:highlight}
\begin{figure*}[!t]
    \centering
    \begin{tabular}{ccc}
        \includegraphics[width=0.27\linewidth]{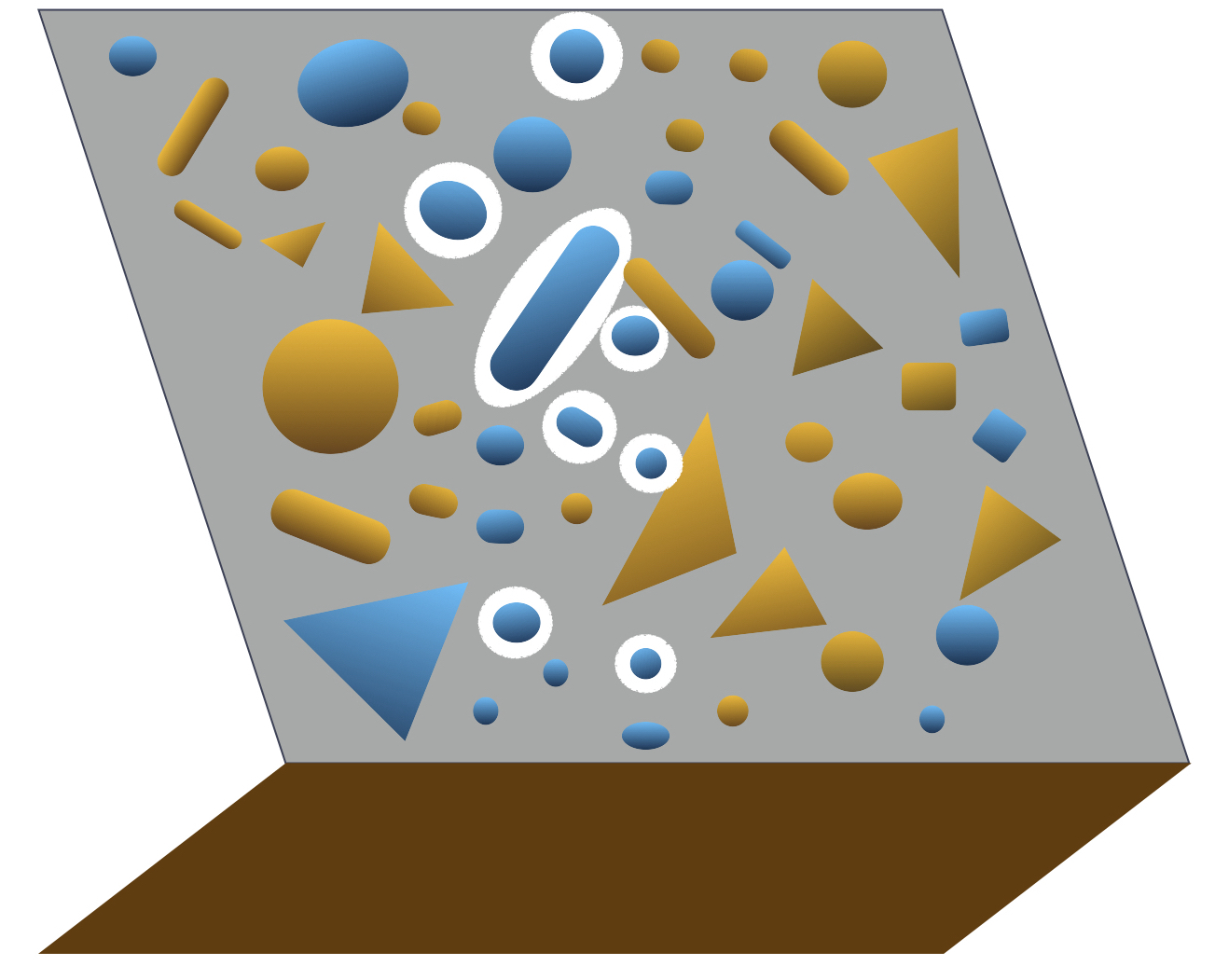}&
        \includegraphics[width=0.27\linewidth]{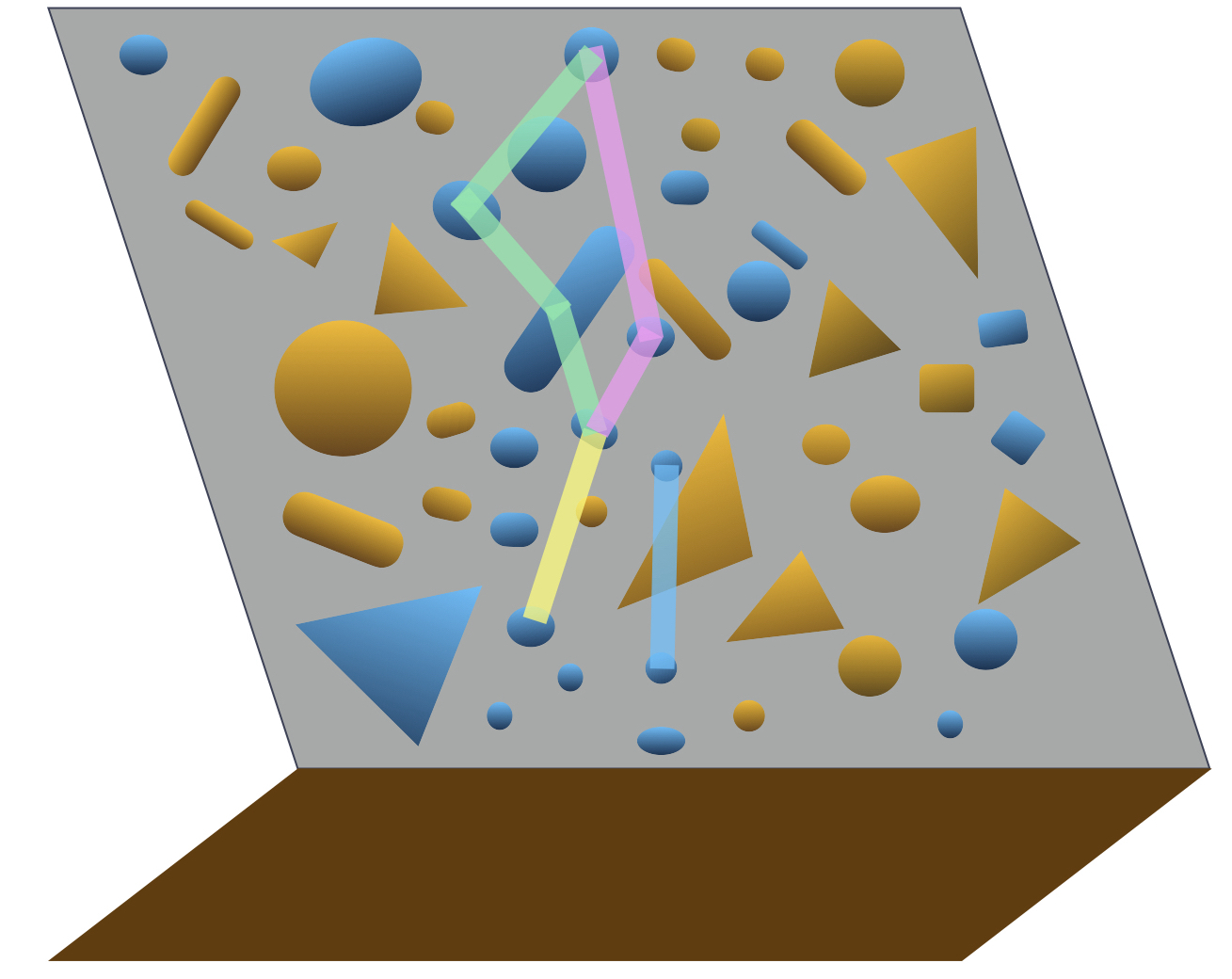}&
        \includegraphics[width=0.27\linewidth]{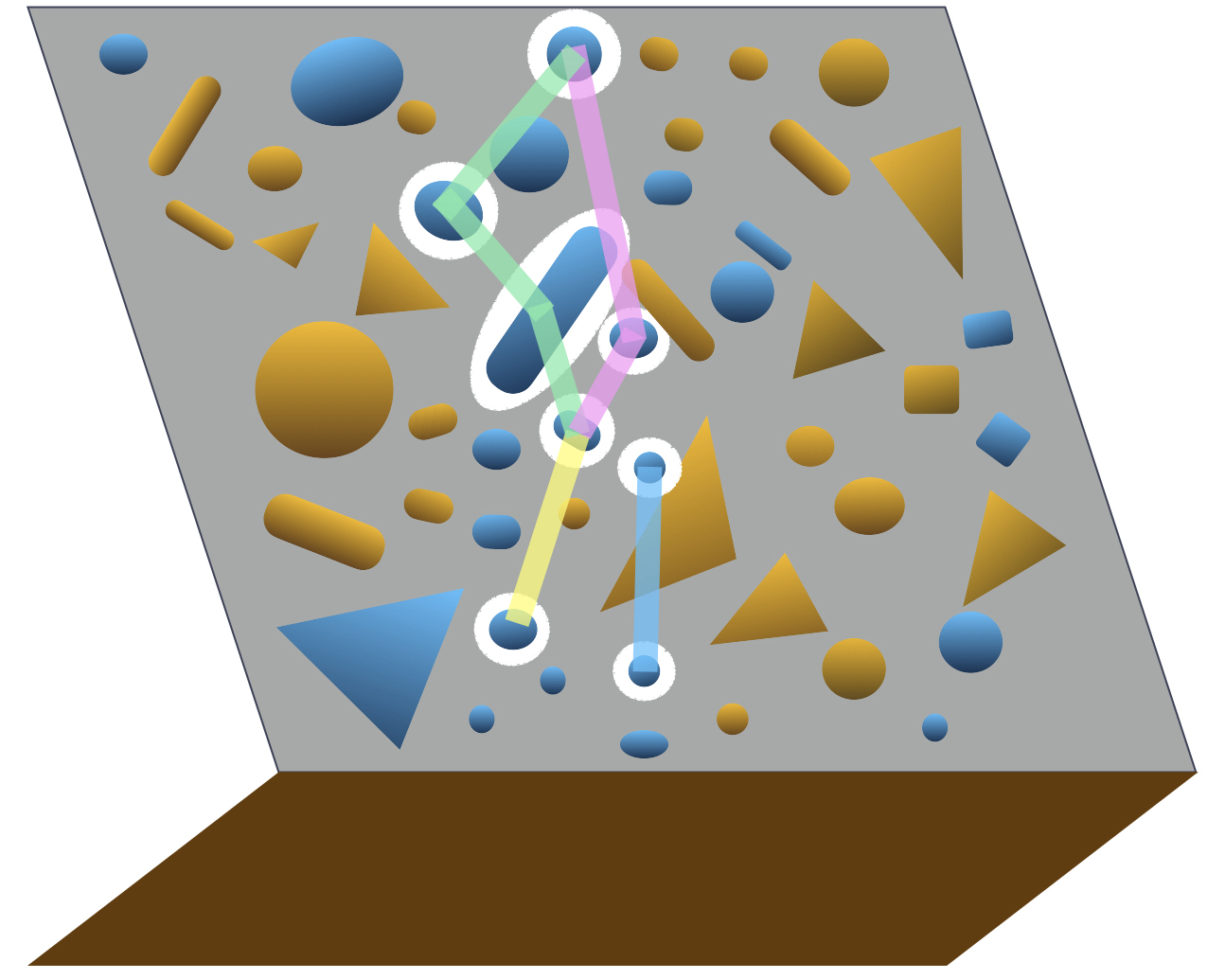}\\
        {\footnotesize (a) Circular}&
        {\footnotesize (b) Linear}&
        {\footnotesize (c) Combined}\\
    \end{tabular}
    \caption{The three projected cues. (a) White circles on the route holds. (b) Lines connecting the route holds, one color per limb. (c) Circles and lines together.}
    \label{fig:highlighting}
\end{figure*}

A climber observing a route has to determine which holds belong to the route, the order in which they are used, which hand or foot takes each hold, when to move, and what kind of movement each move requires, such as matching both hands on one hold or swapping feet. A projected cue can convey only part of this information, and each added piece of information also adds graphics to the wall. We therefore chose three cues that convey different parts of this information (Table~\ref{tab:cueinfo}, Fig.~\ref{fig:highlighting}).

\begin{figure}[tb]
    \centering
    \includegraphics[width=0.9\linewidth]{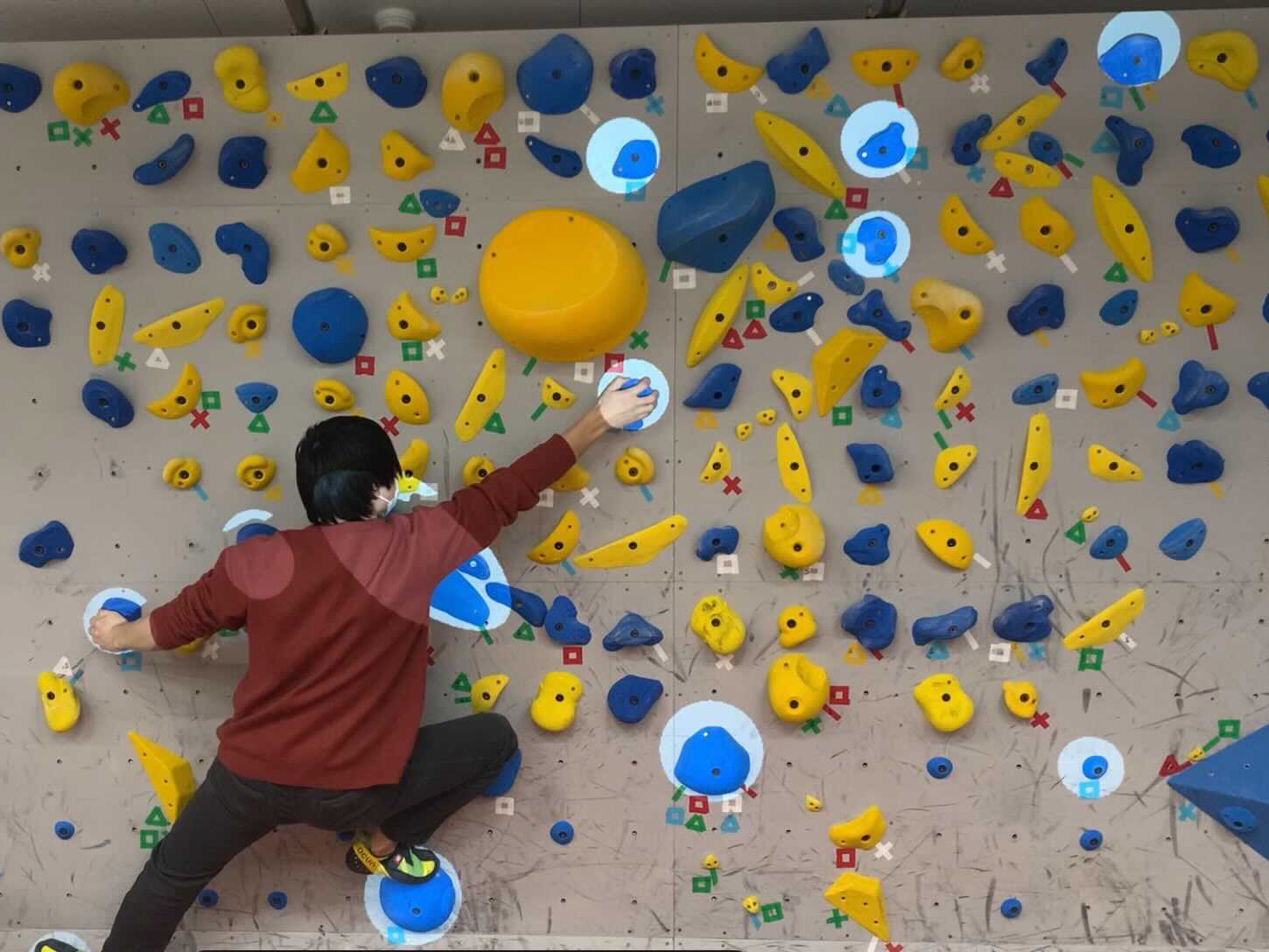}
    \caption{Example of circular highlighting projected onto holds.}
    \label{fig:systemImage}
\end{figure}

\begin{itemize}
    \item \textbf{Circular highlighting} projects a white circle onto each hold of the route (Figs.~\ref{fig:highlighting}(a) and \ref{fig:systemImage}). It conveys which holds belong to the route and where they are, but not their order or the limb for each.
    \item \textbf{Linear highlighting} connects the holds of the route with lines in four colors, one color for each hand and foot: yellow for the left foot, blue for the right foot, green for the left hand, and red for the right hand (Fig.~\ref{fig:highlighting}(b)). The lines are about 50--60\,mm wide, depending on the position on the wall. Each colored line shows the holds taken by one limb in order, running upward from the start of the route (a direction not explained explicitly); the order of moves between limbs is not shown. The holds of the route are indicated only by the ends and bends of the lines; an off-route hold under a line is covered by it in the same way.
    \item \textbf{Combined highlighting} projects both the circles and the lines (Fig.~\ref{fig:highlighting}(c)). It conveys the information of both other cues and has the most graphics.
\end{itemize}

\begin{table}[!t]
\centering
\caption{Information conveyed by each projected cue ($\checkmark$: conveyed; $\triangle$: conveyed only by the ends and bends of the lines; --: not conveyed).}
\label{tab:cueinfo}
\scalebox{0.85}{
\begin{tabular}{lccc}
\hline
Information & Circular & Linear & Combined \\
\hline
Holds belonging to the route & $\checkmark$ & $\triangle$ & $\checkmark$ \\
Order of holds for each limb & -- & $\checkmark$ & $\checkmark$ \\
Limb for each hold & -- & $\checkmark$ & $\checkmark$ \\
Timing of each move & -- & -- & -- \\
Type of movement (e.g., matching) & -- & -- & -- \\
Start and goal holds & -- & -- & -- \\
\hline
\end{tabular}
}
\end{table}

Comparing circular and linear highlighting shows the effect of conveying the order and limb assignment in place of an explicit mark on each hold, and comparing circular and combined highlighting shows the effect of adding the order and limb assignment to the marks on the holds. None of the cues distinguishes the start and goal holds from the other route holds; the tape labels on the wall were the only indication of them. None of the cues conveys the timing or the type of each move, and none changes during the climb; the cues are static and are shown during both observation and climbing. Other designs, such as numbering the holds or showing only the next holds as the climber moves up, were not included in this study and are discussed in Section~\ref{sec:discussion}.

The cues were authored by hand for each route. The experimenter placed a sphere at each route hold in the Unity scene for circular highlighting and cylinders between consecutive holds of each limb for linear highlighting, in colors chosen to remain distinguishable when projected. The sequence of holds for each limb was determined by the authors, including experienced climbers, from the sequence intended by the route setter. Automatic detection of holds and estimation of the limb sequence are outside the scope of this study.

\section{User Study}
\label{sec:user_study}
We conducted a user study to compare how the three projected cues and a condition without projection affect route observation and climbing. The study was approved by the Institutional Review Board of the Cybermedia Center, Osaka University (approval No.~\mbox{2021-13}), and all participants gave written informed consent.

\subsection{Design}
\label{sec:design}
The study had one within-participant factor, the cue, with four levels: no projection, circular highlighting, linear highlighting, and combined highlighting. Each participant climbed four routes (A--D), each under a different cue. The routes were climbed in one of four cyclic orders (A-B-C-D, B-C-D-A, C-D-A-B, or D-A-B-C) and the cues were presented in one of four cyclic orders of no projection, circular, linear, and combined highlighting, so that both orders formed Latin squares; crossing the two gave 16 sequences. Participants were assigned to the sequences in the order in which they took part, without randomization. Each sequence was used by six participants, so that each route was climbed 24 times under each cue.

\subsection{Environment and Routes}
\begin{figure*}[!t]
\centering
    \includegraphics[width=0.75\linewidth]{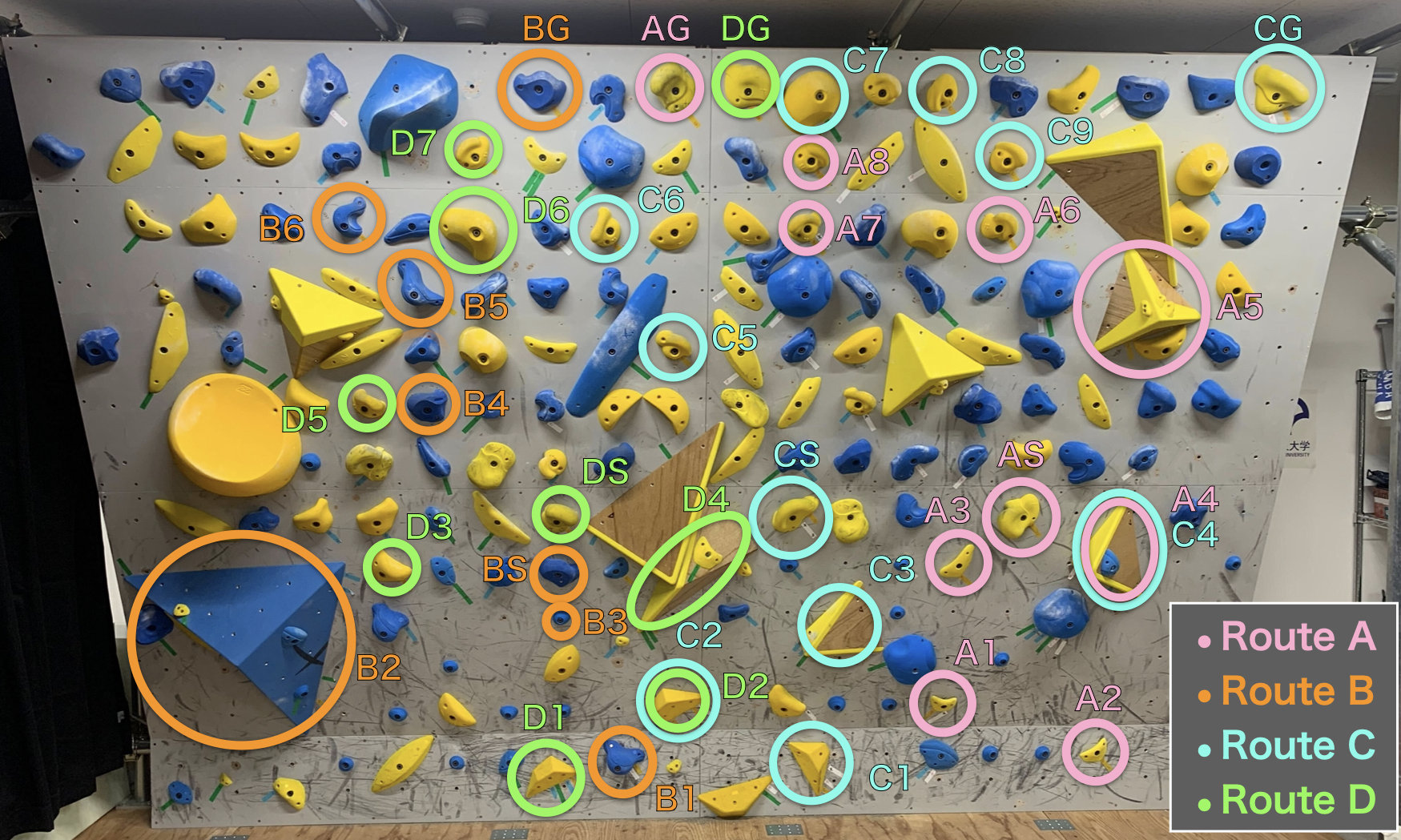}
    \caption{The four experimental routes on the bouldering wall used in the study. \textit{AS} and \textit{AG} denote the start and goal holds of Route A, and \textit{A1}, \textit{A2}, \ldots\ denote its other holds; Routes B--D are labeled in the same way. All routes are approximately V0 on the V-scale.}
    \label{fig:routes}
\end{figure*}

The study was conducted on a bouldering wall 4.1\,m wide and 2.5\,m high, inclined 20 degrees past vertical. An experienced route setter set four routes of comparable difficulty, all approximately V0 on the V-scale (Fig.~\ref{fig:routes}), and a separate tutorial route. The route setter was asked to set routes on which participants with different skill levels would reach different heights and whose difficulty depended on finding the right movements rather than on strength. Each hold of an experimental route was marked with a 15-mm-wide strip of yellow tape bearing the letter of the route, placed beside or below the hold; the start and goal holds were marked with two strips labeled S and G. The tape labels were deliberately made somewhat hard to follow, as they are in many commercial gyms, and the holds and tape of 11 other routes on the same wall surrounded the experimental routes.

\begin{figure}[!t]
    \centering
    \includegraphics[width=0.9\linewidth]{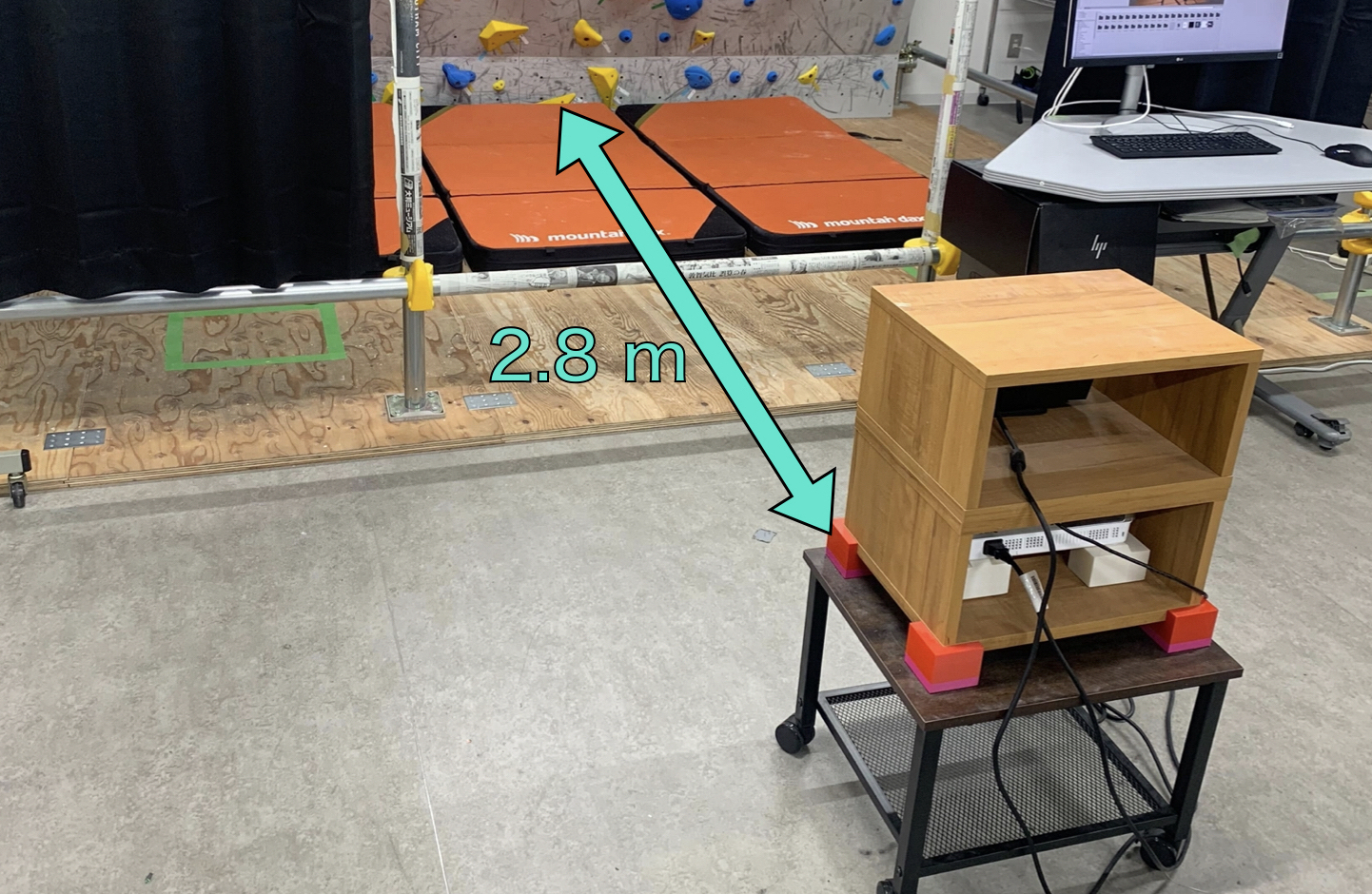}
    \caption{Placement of the projector-camera system, approximately 2.8\,m from the wall.}
    \label{fig:projectorPosition}
\end{figure}

The projector-camera system was placed approximately 2.8\,m from the base of the wall (Fig.~\ref{fig:projectorPosition}). The room lighting was dimmed so that the projected cues were visible while the wall remained safe to climb; the illuminance on the wall was 84.1\,lx.

\subsection{Participants and Procedure}
Ninety-six participants aged 18--27 (65 male, 31 female; heights 146--183\,cm) took part from December 2021 to January 2022. For 61 participants, this was their first bouldering session; 13, 7, and 15 were in their second, third, and fourth or later sessions, respectively.

Participants were first given an explanation of the basic rules of bouldering, of route observation, and of the projected cues. They were told not to try to climb fast, and the experimenter did not answer questions about how to climb the routes or where their holds were. After climbing the tutorial route, they performed four trials, one on each of Routes A--D. Each trial lasted 3\,min, within which participants could observe the route and attempt it as many times as they wished. Only the holds of the route could be used, for both the hands and the feet. An attempt started with both hands on the single start hold and both feet off the ground in a stable position, and the route was completed when the participant held the goal hold with both hands. For the observation before the first attempt, participants stood at a position from which the projected cues were fully visible. The trial ended when the route was completed or when 3\,min had elapsed, after which participants rested and answered a questionnaire about the trial; after the four trials, they answered a post-experiment questionnaire. Crash mats covered the floor, and all trials were recorded on video.

\subsection{Measures}
\label{sec:measures}
The following measures were coded from the videos by one author (T.~H.).
\begin{itemize}
    \item \textbf{Time before the first attempt}: the time from the start of the trial until the first attempt began. Some participants tried individual moves by climbing onto the route from a hold other than the start hold; this time was counted as part of the time before the first attempt. The measure therefore includes observation from the ground and, for some participants, trying moves on the wall; for brevity, we refer to it as observation time. Trials in which no attempt was begun (17 trials) were assigned 180\,s.
    \item \textbf{Number of attempts}: the number of attempts within the trial.
    \item \textbf{Number of moves}: the number of hand moves from the start hold to other holds of the route in the attempt with the most such moves, and 0 in trials without an attempt. Because some participants skipped holds, completions took 2--6 moves, depending on the route.
    \item \textbf{Use of off-route holds}: whether the participant used at least one hold that did not belong to the route, including volumes, with a hand or a foot. For 7 of the 17 trials without an attempt, the entry was blank and was treated as no use.
    \item \textbf{Completion}: whether the participant held the goal hold with both hands in an attempt that used only route holds. Completions recorded in trials with a single attempt in which an off-route hold was used were not counted (6 trials). In 4 trials with several attempts, an off-route hold was used and the route was completed, and the records do not show in which attempt the off-route hold was used; these completions were counted.
\end{itemize}
Immediately after each trial, participants rated the items in Table~\ref{tab:questionnaire1} on a five-point Likert scale (1: strongly disagree; 5: strongly agree); Q1-5 to Q1-8, which refer to the projection, were asked only after trials with a projected cue. The post-experiment questionnaire (Table~\ref{tab:questionnaire2}) asked about the room lighting (Q2-1) and asked participants to rank the conditions by preference (Q2-2) and the routes by difficulty (Q2-3).

\begin{table}[!b]
\caption{Questionnaire items answered immediately after each trial. Q1-5 to Q1-8 were answered only after trials with a projected cue.}
\label{tab:questionnaire1}
\centering
\scalebox{0.75}{
\begin{tabular}{ll}
\hline
Q1-1 & It was easy to identify the holds during observation. \\
Q1-2 & It was easy to identify the next hold while climbing. \\
Q1-3 & I did not lose sight of the holds while climbing. \\
Q1-4 & It was easy to see how to climb the route. \\
Q1-5 & The projection made the holds easier to see. \\
Q1-6 & The colors of the projection were easy to see. \\
Q1-7 & I felt that the projection made the climb easier. \\
Q1-8 & I would like to use the projection in the future. \\
\hline
\end{tabular}
}
\end{table}

\begin{table}[!b]
\caption{Post-experiment questionnaire items.}
\label{tab:questionnaire2}
\centering
\scalebox{0.75}{
\begin{tabular}{ll}
\hline
Q2-1 & The darkness of the room did not bother me. \\
Q2-2 & Rank the four conditions from best to worst. \\
Q2-3 & Rank the four routes from most to least difficult. \\
\hline
\end{tabular}
}
\end{table}

\subsection{Data Preparation and Analysis}
\label{sec:analysis}
The video recordings were coded on sheets organized by route and cue, in which the trials of each route--cue combination were listed in the order in which participants took part, without participant identifiers. We assigned the $k$-th record of each combination to the $k$-th participant who climbed that combination, using the blank columns that an earlier version of the sheets kept at the positions of missing trials. Of the 384 trials, 365 had a record. We checked the assignment in two ways. The number of moves was consistent within participants (intraclass correlation 0.50, against 0.00 on average when records were reassigned at random within each combination), and the bouldering experience reported in the questionnaire, which carries the participant's identifier and was not used for the assignment, was associated with completion under this assignment (Spearman's $r = 0.22$, against a 99th percentile of 0.10 under random reassignment).

The behavioral measures were analyzed with mixed-effects models that included a random intercept for each participant and fixed effects of cue, route, trial position, and bouldering experience (first, second or third, or fourth or later session): logistic models for completion and off-route holds, Poisson models for attempts and moves, and a linear model for the logarithm of observation time. The effect of cue was tested with a likelihood-ratio test, and the six pairwise comparisons were corrected with Holm's method. Effects are reported as ratios with 95\% confidence intervals adjusted for the six pairs (Bonferroni); because Holm's method is less conservative than Bonferroni's, an interval can include 1 for a comparison that is significant. Observation time was measured as planned but was not the planned main outcome.

Items Q1-1 to Q1-4 were analyzed with cumulative link mixed models with the same effects. Items Q1-5 to Q1-8 (for the three cues, excluding 40 item responses from 28 participants after trials without projection) and the rankings Q2-2 and Q2-3 were analyzed with Friedman tests followed by Wilcoxon signed-rank tests with Holm's correction. Analyses used R 4.4.3 with lme4, ordinal, emmeans, geepack, and survival.

\subsection{Results}
\label{sec:results}
Table~\ref{tab:descriptive} summarizes the behavioral measures for each cue, and Table~\ref{tab:models} shows the pairwise comparisons estimated by the mixed-effects models.

\begin{table}[!t]
\centering
\caption{Behavioral measures for each cue (365 trials of 94 participants). Observation time is the model-estimated geometric mean, with 180\,s for trials without an attempt; other values are counts and means.}
\label{tab:descriptive}
\scalebox{0.78}{
\begin{tabular}{lcccc}
\hline
 & None & Circular & Linear & Combined \\
\hline
Trials & 91 & 91 & 91 & 92 \\
Observation time [s] & 82.5 & 38.8 & 51.2 & 47.8 \\
No attempt begun & 6/90 & 2/91 & 3/91 & 6/91 \\
Attempts per trial (mean) & 3.0 & 4.8 & 4.0 & 4.6 \\
Moves (mean) & 1.22 & 1.91 & 1.63 & 1.74 \\
Completed & 5/91 & 17/91 & 11/91 & 13/92 \\
Off-route holds used & 14/90 & 6/91 & 26/91 & 15/91 \\
\hline
\end{tabular}
}
\end{table}

\begin{table*}[!t]
\centering
\caption{Pairwise comparisons between cues estimated by the mixed-effects models: ratios of the first to the second cue with Bonferroni-adjusted 95\% confidence intervals (observation time: ratio of geometric means; attempts and moves: rate ratio; completion and off-route holds: odds ratio). Holm-corrected *: $p<.05$; **: $p<.01$; ***: $p<.001$.}
\label{tab:models}
\scalebox{0.8}{
\begin{tabular}{lccccc}
\hline
Comparison & Observation time & Attempts & Moves & Completion & Off-route holds \\
\hline
Circular / None & 0.47 [0.39, 0.57]*** & 1.60 [1.31, 1.97]*** & 1.62 [1.18, 2.23]*** & 12.1 [1.4, 105]* & 0.30 [0.06, 1.39] \\
Linear / None & 0.62 [0.51, 0.75]*** & 1.35 [1.09, 1.66]*** & 1.36 [0.98, 1.89] & 3.9 [0.5, 32.6] & 2.78 [0.88, 8.79] \\
Combined / None & 0.58 [0.48, 0.70]*** & 1.53 [1.24, 1.88]*** & 1.45 [1.05, 2.01]* & 7.4 [0.8, 66.3] & 1.13 [0.33, 3.89] \\
Linear / Circular & 1.32 [1.09, 1.59]*** & 0.84 [0.70, 1.01]* & 0.84 [0.62, 1.12] & 0.32 [0.06, 1.81] & 9.3 [2.1, 42.5]*** \\
Combined / Circular & 1.23 [1.02, 1.49]** & 0.95 [0.80, 1.14] & 0.89 [0.67, 1.19] & 0.61 [0.12, 3.17] & 3.80 [0.81, 17.8] \\
Combined / Linear & 0.93 [0.77, 1.13] & 1.14 [0.94, 1.37] & 1.07 [0.79, 1.44] & 1.9 [0.3, 11.7] & 0.41 [0.13, 1.27] \\
\hline
Cue effect, $\chi^2(3)$ & 101.9 ($p<.001$) & 44.4 ($p<.001$) & 16.8 ($p<.001$) & 12.7 ($p=.005$) & 20.3 ($p<.001$) \\
Trials in the model & 363 & 363 & 364 & 365 & 363 \\
\hline
\end{tabular}
}
\end{table*}

\subsubsection{Observation Time and Attempts}
\begin{figure}[!t]
    \centering
    \includegraphics[width=\linewidth]{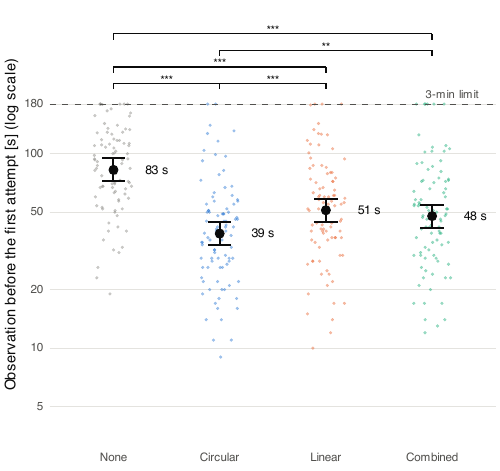}
    \caption{Observation time before the first attempt (log scale). Small dots: 363 trials, those without an attempt at 180\,s. Black dots and bars: model-estimated geometric means and 95\% confidence intervals. Brackets: Holm-corrected comparisons (**: $p<.01$; ***: $p<.001$).}
    \label{fig:observation}
\end{figure}

All three cues shortened the observation before the first attempt (Fig.~\ref{fig:observation}). The geometric mean was 82.5\,s without projection and 38.8\,s with circular highlighting, a ratio of 0.47; the ratios were 0.62 with linear and 0.58 with combined highlighting (all $p<.001$). Observation was shorter with circular highlighting than with linear ($p<.001$) or combined highlighting ($p=.008$). Adding the interaction between cue and experience (first session or not) did not improve the model ($\chi^2(3) = 1.39$, $p = .71$; ratio for circular highlighting 0.48 in the first session and 0.44 otherwise). Participants also made more attempts within the trial with every cue than without projection (all $p<.001$), and fewer with linear than with circular highlighting ($p=.039$).

\subsubsection{Moves and Completion}
The number of moves was larger with circular (rate ratio 1.62, $p<.001$) and combined highlighting (1.45, $p=.013$) than without projection, and did not differ between the three cues. Of the 365 trials, 46 were completed. The cue had an effect on completion ($\chi^2(3) = 12.7$, $p=.005$), and only circular highlighting differed from no projection (odds ratio 12.1, 95\% confidence interval 1.4--105, $p=.014$). The median time from the start of the trial to completion was 100, 70, 90, and 97\,s without projection and with the circular, linear, and combined cues (45 completed trials with recorded times; not tested).

\subsubsection{Use of Off-route Holds}
\begin{figure}[!t]
    \centering
    \begin{tabular}{cc}
        \includegraphics[width=0.45\linewidth]{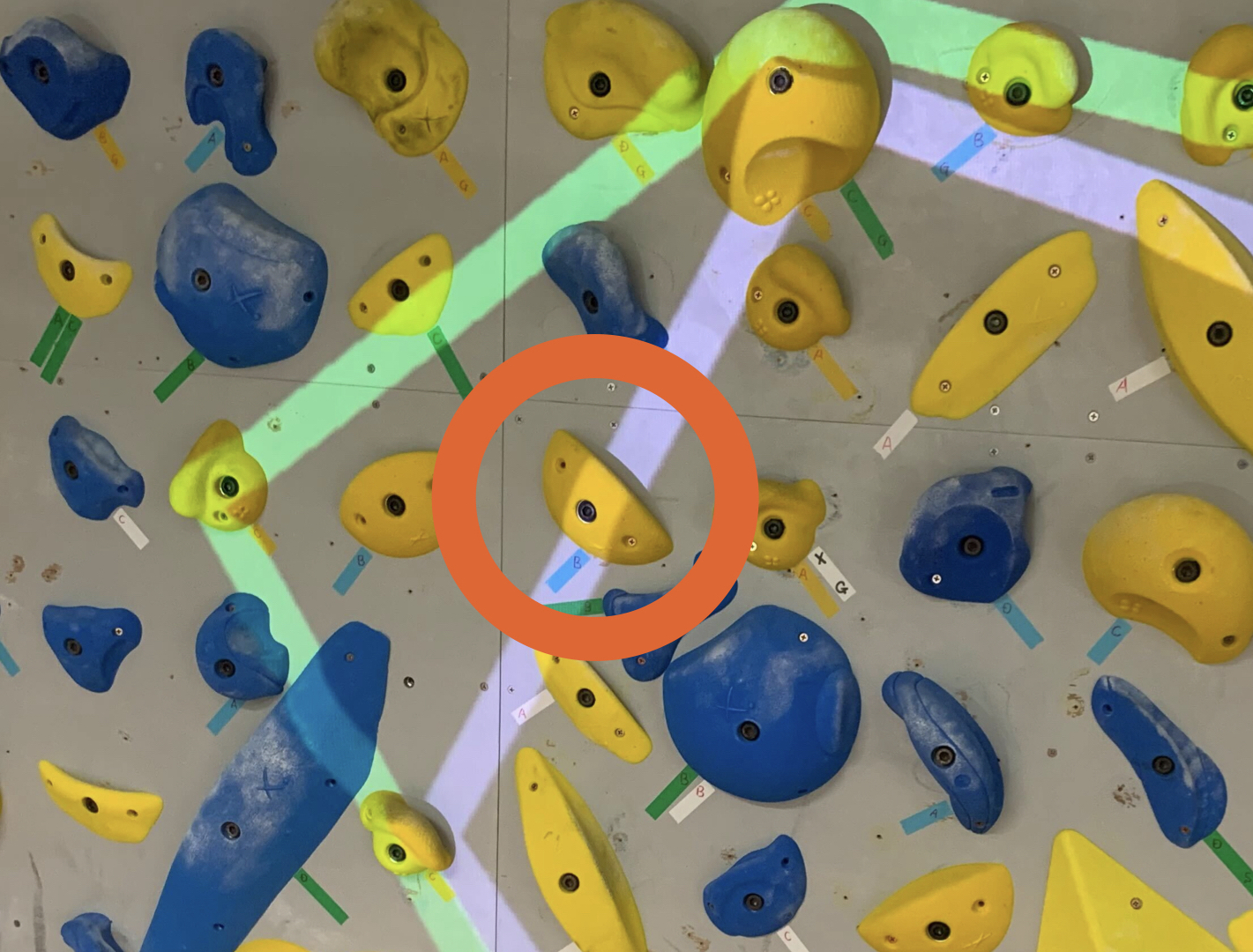}&
        \includegraphics[width=0.45\linewidth]{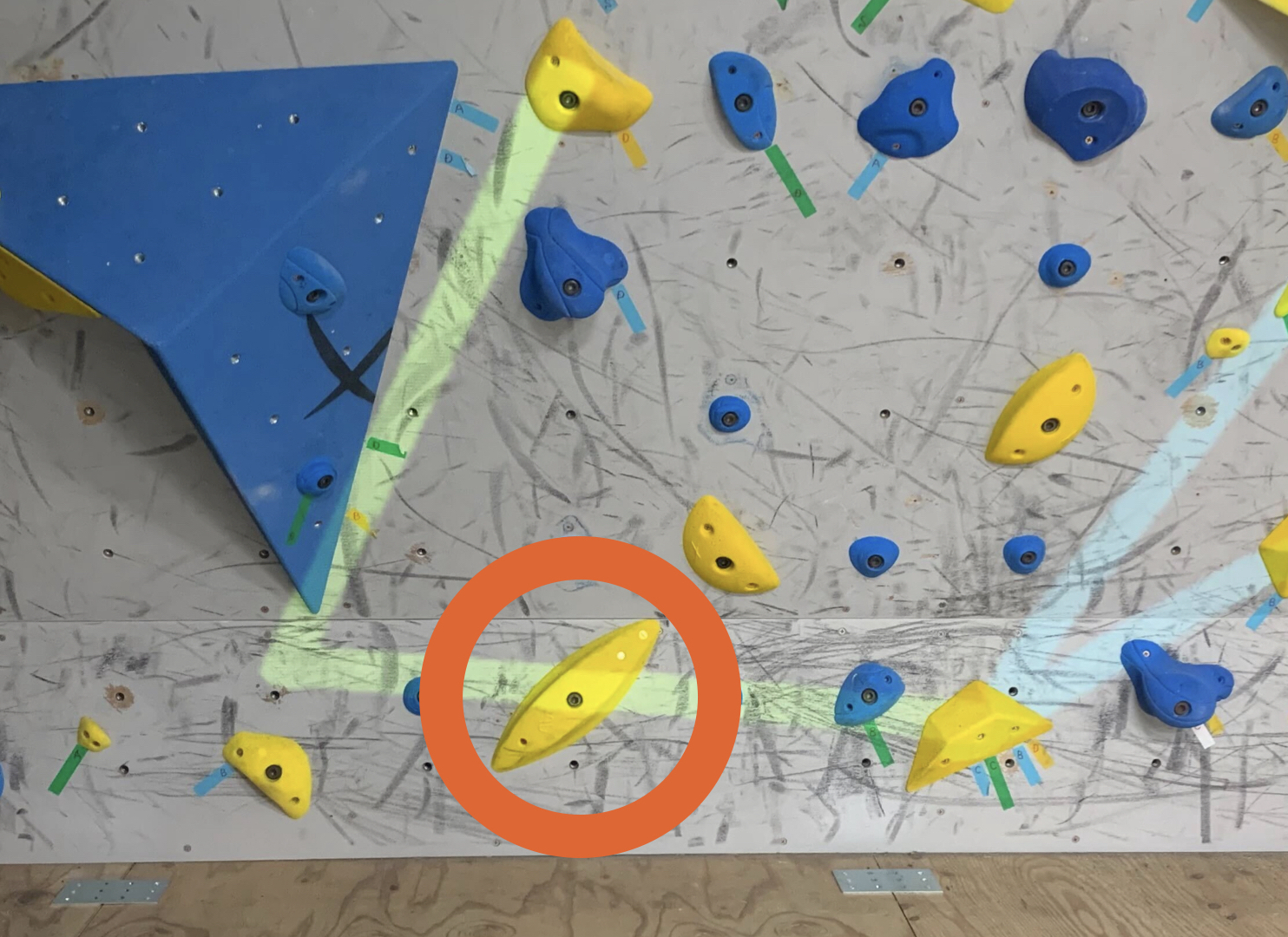}\\
        {\footnotesize (a) Route C}&
        {\footnotesize (b) Route D}\\
    \end{tabular}
    \caption{Holds that do not belong to the route lying under the projected lines on Routes C and D.}
    \label{fig:holdOnLine}
\end{figure}

The odds of using an off-route hold were 9.3 times higher with linear than with circular highlighting ($p<.001$; 26 vs.\ 6 of 91 trials), and no other pair differed significantly. In an exploratory comparison by route, the use of off-route holds with linear highlighting was concentrated on Routes C (7 of 23 trials) and D (11 of 22 trials), where it exceeded that in the other three conditions (Fisher's exact test, Holm-corrected over the four routes, $p=.018$ and $p=.049$). On these routes, holds that did not belong to the route lay under the projected lines (Fig.~\ref{fig:holdOnLine}).

\subsubsection{Sensitivity Analyses}
\label{sec:robustness}
Table~\ref{tab:sensitivity} shows the main comparisons under alternative analyses: generalized estimating equations (GEE) instead of mixed-effects models; an alternative assignment of records in which each missing trial was placed where it maximized the consistency of performance; for observation time, 300 assignments with the positions of the missing trials in each route--cue combination drawn at random, a log-normal regression with robust standard errors that treated trials without an attempt as censored at 180\,s, and the first trials only, which have no preceding cue; and other codings of the four uncertain completions (counted as not completed), the seven blank off-route entries (treated as missing), and two trials in which the participant may have believed the feet were free (excluded). All comparisons that were significant in the main analysis remained significant, including in all 300 random assignments (largest $p=.029$), except for the differences between the cues in the first trials (91 trials), which had the same direction but less precision.

\begin{table}[!t]
\centering
\caption{Main comparisons under the sensitivity analyses (ratios as in Table~\ref{tab:models}; Holm-corrected *: $p<.05$; **: $p<.01$; ***: $p<.001$).}
\label{tab:sensitivity}
\setlength{\tabcolsep}{3pt}
\scalebox{0.72}{
\begin{tabular}{lccccc}
\hline
 & \multicolumn{3}{c}{Observation time} & Completion & Off-route \\
Analysis & \shortstack{Circ.\\/ None} & \shortstack{Lin.\\/ Circ.} & \shortstack{Comb.\\/ Circ.} & \shortstack{Circ.\\/ None} & \shortstack{Lin.\\/ Circ.} \\
\hline
Main analysis & 0.47*** & 1.32*** & 1.23** & 12.1* & 9.3*** \\
GEE & 0.47*** & 1.32*** & 1.23* & 4.6* & 6.3*** \\
Alternative assignment & 0.47*** & 1.32** & 1.24* & 14.9* & 10.8*** \\
Random assignments (min--max) & 0.46--0.48 & 1.29--1.34 & 1.22--1.26 & -- & -- \\
Censored at 180\,s & 0.46*** & 1.33*** & 1.26** & -- & -- \\
First trials only (91) & 0.42*** & 1.61 & 1.46 & -- & -- \\
Uncertain completions & -- & -- & -- & 11.2* & -- \\
Blank off-route entries & -- & -- & -- & -- & 9.1*** \\
Feet-free trials excluded & -- & -- & -- & 12.3* & 9.9*** \\
\hline
\end{tabular}
}
\end{table}

\subsubsection{Questionnaires}
\begin{figure*}[!t]
    \centering
    \includegraphics[width=\linewidth]{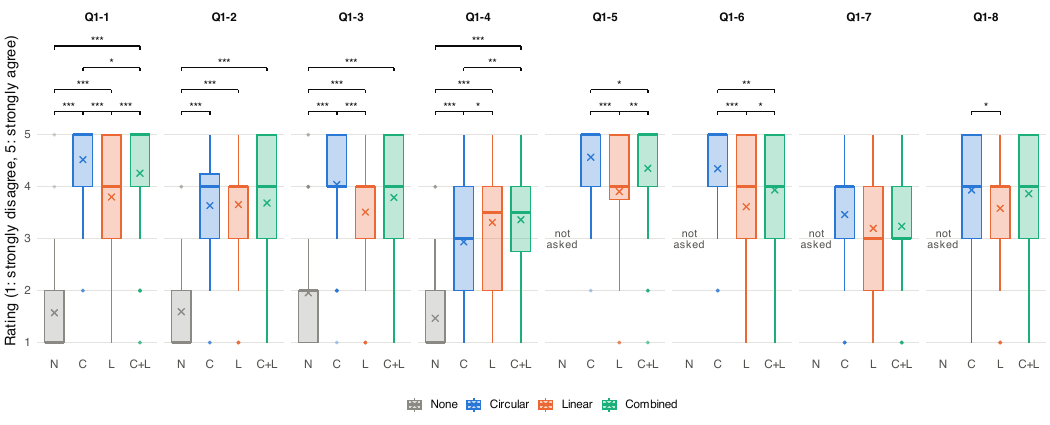}
    \caption{Ratings after each trial (N: no projection; C: circular; L: linear; C+L: combined). Boxes: quartiles; thick lines: medians; crosses: means. Q1-5 to Q1-8 were asked only after trials with a cue. Brackets: Holm-corrected comparisons (*: $p<.05$; **: $p<.01$; ***: $p<.001$).}
    \label{fig:result1}
\end{figure*}

The responses to the items answered after each trial are shown in Fig.~\ref{fig:result1}. For each of Q1-1 to Q1-4, ratings with every cue were higher than without projection ($p<.001$). Among the cues, circular highlighting was rated higher than linear ($p<.001$) and combined highlighting ($p=.043$) for identifying holds during observation (Q1-1) and higher than linear highlighting for not losing sight of holds (Q1-3, $p<.001$), whereas linear ($p=.012$) and combined highlighting ($p=.007$) were rated higher than circular highlighting for seeing how to climb (Q1-4). The cues did not differ in identifying the next hold while climbing (Q1-2). For the items about the projection (95 participants), circular highlighting was rated highest for making holds easier to see (Q1-5) and for the visibility of the colors (Q1-6; all pairs differed, $p\le.029$), and higher than linear highlighting for future use (Q1-8, $p=.021$); the cues did not differ in whether climbing felt easier (Q1-7).

\begin{figure}[!t]
    \centering
    \includegraphics[width=\linewidth]{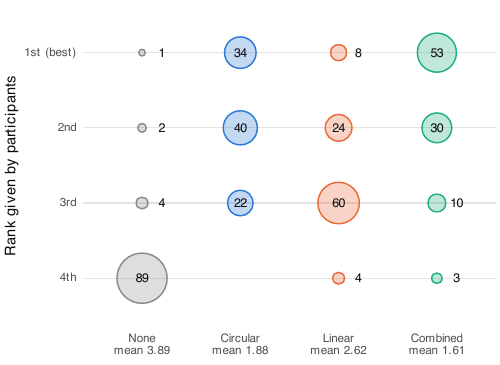}
    \caption{Preference ranking of the four conditions (Q2-2; 96 participants). Circle areas and numbers give the number of participants who gave each rank; mean ranks are shown under the condition names. All pairs differed (Holm-corrected).}
    \label{fig:effect rank}
\end{figure}

After the trials, 88 of 95 respondents agreed that the darkness of the room did not bother them (Q2-1). In the preference ranking (Q2-2, Fig.~\ref{fig:effect rank}), the mean ranks were 1.61 for combined, 1.88 for circular, 2.62 for linear highlighting, and 3.89 for no projection, and all pairs differed (Friedman $\chi^2(3) = 179.1$, $p<.001$; combined vs.\ circular, $p=.044$; all other pairs, $p<.001$). Route A was ranked as the most difficult route (Q2-3) and differed from the other three ($p\le.026$).

\subsection{Discussion}
\label{sec:discussion}

\subsubsection{Observation, Attempts, and Completion}
The clearest effect of the projected cues was the shorter time before the first attempt, roughly halved with circular highlighting. Without projection, participants had to find the route holds among those of 11 other routes, and circular highlighting marked these holds directly. Because all participants faced an unfamiliar wall and the shortening did not differ detectably between those in their first session and the others, it is consistent with the cues mainly shortening the search for holds; without gaze measurements, however, we cannot separate finding holds from planning movements and trying moves on the wall. Observation was longer with linear and combined highlighting than with circular highlighting, although they convey more information (Table~\ref{tab:cueinfo}): participants may have spent time reading the order and limb assignment, or the route holds may have been harder to identify from lines, as the lower ratings of linear highlighting for Q1-1 and Q1-5 suggest.

The time saved was followed by more attempts. Because moves were counted in the attempt with the most moves, more attempts gave more chances to progress, so the attempts and moves describe how participants used the trial rather than an improvement in climbing skill. Completion was rare, and most of its variation was between participants; circular highlighting increased it in every analysis, by an uncertain amount. Sanchez et al.\ found that route preview changed climbers' movements during an ascent but not route completion \cite{sanchez2012efficacy}; in our study as well, the clearest effect was on how soon participants started climbing, not on completion.

\subsubsection{Lines and Off-route Holds}
The more frequent use of off-route holds with linear highlighting concerns the present line design rather than the addition of order information as such: the holds of the route are indicated only by the ends and bends of the lines (Table~\ref{tab:cueinfo}), and a hold under a line is covered in the same way as a hold at its end. Participants who used the lines to find holds could therefore take a covered hold for a route hold, which agrees with the concentration of off-route holds on Routes C and D. This explanation was not tested directly, because the holds used were not identified and the routes also differed in other respects.

\subsubsection{Preference and Performance}
Participants preferred combined to circular highlighting and rated the cues with lines higher for seeing how to climb (Q1-4), but combined highlighting led to longer observation than circular highlighting, and no behavioral advantage of it was detected.

\subsubsection{Design Recommendations}
The first and last of the following recommendations follow from the results; the second and third are proposals derived from them that remain to be evaluated.
\begin{itemize}
    \item \textbf{Mark each route hold directly.} Of the three cues, a mark on each hold shortened observation the most and had the fewest trials with off-route holds.
    \item \textbf{Do not let lines pass over holds that are not part of the route.} Lines that show the order or limb assignment should be routed around off-route holds or end in a mark that distinguishes route holds.
    \item \textbf{Show only the holds that are needed next.} As the depth sensor can track the climber, the cues could show only the next holds of each limb, reducing the graphics and the holds under them.
    \item \textbf{Evaluate cues by behavior as well as preference.} The most preferred cue was not the most effective one in the behavioral measures.
\end{itemize}

\subsubsection{Limitations}
First, the single projector behind the climber let the climber's body cast shadows on the cues during climbing, which several participants reported; its effect on the comparison during climbing was not measured, whereas the observation before the first attempt was unaffected. Shadows can be reduced with several projectors that compensate for occluded regions \cite{sukthankar2001dynamic,summet2007shadow} or with optical designs that project without shadows \cite{hiratani2023shadowless}.

Second, the room was dimmed to 84.1\,lx, and the tape labels were deliberately hard to follow, so the results show how the cues compare with each other and with such labels, not with the clearly marked routes and bright lighting of many commercial gyms. Kajastila and H\"am\"al\"ainen reported that 5000-lm projectors were sufficient for a 3 $\times$ 4\,m wall \cite{augmentedWall}, but the brightness needed also depends on the ambient light and the wall. The cues also differed simultaneously in information, color, and projected area, so the results concern these implementations rather than any single factor.

Third, the study used one wall, four easy routes, and mostly novice participants in a single session; the results may differ for harder routes, experienced climbers, and repeated use. Because the cue orders were cyclic, each cue was always preceded by the same cue, and in the second to fourth trials an effect carried over from the preceding cue is confounded with that of the cue itself.

Fourth, the records were assigned to participants by their order on the coding sheets, the identity of the 19 missing trials could not be verified, and the videos were coded by a single coder who could see the condition. The sensitivity analyses limit, but do not remove, the uncertainty from these sources.

Fifth, the cues were authored by hand; for frequently reset routes, reducing this work, for example by detecting route holds automatically, is a practical issue.

\section{Conclusion}
\label{sec:conclusion}
We compared three cues projected onto a bouldering wall (circles on the route holds, limb-colored lines connecting the holds, and both combined) with a condition without projection in a study with 96 participants. All three cues shortened the time before the first attempt; with circles, its geometric mean fell from 83\,s to 39\,s. Participants then made more attempts within the trial and, with circles and the combined cue, more moves. Completion was higher with circles than without projection, although the size of the effect is uncertain. Lines led to more use of off-route holds than circles, mainly on routes where such holds lay under the lines. Participants preferred the combined cue, although no behavioral advantage of it over circles was detected.

These results indicate that, under the conditions of this study, a mark on each route hold was the most effective of the three cues in shortening the time before the first attempt. They also suggest that lines conveying the order and limb assignment of holds should not pass over other holds. Future work includes revised line designs, cues that show only the next holds as the climber progresses, multiple projectors to reduce the shadows cast by the climber, automatic generation of the cues from camera images, and studies of whether climbers who practice with the cues learn to observe routes without them.

\section*{Declaration of Generative AI Use}
During the preparation of this work, the authors used Claude (Anthropic) to assist in revising the text, writing the scripts for the statistical analysis and the figures, and checking the descriptions of the cited works against their sources. The authors reviewed and edited all output and take full responsibility for the article.

\bibliographystyle{ite_mta}
\bibliography{reference}

\begin{biography}
\profile{Yuki}[]{Uranishi}[]{received the Ph.D. degree from the Graduate School of Information Science, Nara Institute of Science and Technology, Japan, in 2008. He is a Professor at the D3 Center, the University of Osaka, Japan. Before joining the University of Osaka, he was a Research Fellow with the Japan Society for the Promotion of Science from 2008 to 2009, an Assistant Professor at Nara Institute of Science and Technology from 2009 to 2012, at Osaka University from 2012 to 2014, and at Kyoto University Hospital from 2014 to 2016. He was an Associate Professor at the Cybermedia Center at Osaka University from 2016 to 2024. His research interests include computer vision, XR, and human-computer interaction.}
\profile{Tamon}[]{Hamamoto}[]{received his B.E. and M.E. degrees from Osaka University, Japan, in 2020 and 2022, respectively. His research interest is in augmented reality.}
\profile{Chang}[]{Liu}[]{received his Ph.D. in Information Science from the Graduate School of Information Science and Technology, Osaka University, in 2020, and is currently a program-specific senior lecturer at the Graduate School of Medicine, Kyoto University. His research interests include human-computer interaction and medical VR/AR applications.}
\profile{Goshiro}[]{Yamamoto}[]{received his Ph.D. in engineering from Osaka University. He is currently a program-specific professor at Kyoto University Hospital. His main research interests are human-computer interaction, digital transformation, and medical informatics.}
\profile{Masato}[]{Kobayashi}[]{received the Bachelor of Maritime Sciences, Master of Maritime Sciences, and Doctor of Engineering degrees from Kobe University, Japan, in 2017, 2019, and 2022. From 2019 to 2021, he was with Seiko Epson Corporation, Japan. From 2021 to 2022, he worked as a research intern at OMRON SINIC X Corporation, Japan. Since 2022, he has been an Academic Researcher at Kobe University, Japan. Since 2023, he has been with the University of Osaka, Japan, where he is currently an Assistant Professor. His current research interests include robot learning, motion control, XR, HCI, and AI.}
\profile{Photchara}[]{Ratsamee}[]{received the M.E. and Ph.D. degrees from the Graduate School of Engineering Science, Osaka University, Japan, in 2012 and 2015, respectively. From 2015 to 2022, he served as an Assistant Professor at the Cybermedia Center, Osaka University. From 2022 to 2024, he was an Associate Professor (Lecturer) at the Faculty of Robotics and Design, Osaka Institute of Technology. He is currently an Associate Professor at the Faculty of Engineering Science, Department of Mechanical Engineering, Kansai University. His research interests include robot vision, rescue robotics, human--robot interaction, and haptics in mixed reality.}
\profile{Haruo}[]{Takemura}[]{received the B.E., M.E., and Ph.D. degrees from Osaka University in 1982, 1984, and 1987, respectively. In 1987, he joined the Advanced Telecommunication Research Institute, International. In 1994, he joined Nara Institute of Science and Technology as an Associate Professor at the Graduate School of Information Science and Technology. From 1998 to 1999, he was a Visiting Associate Professor at the University of Toronto, Ontario, Canada. He was a professor at the Cybermedia Center, the University of Osaka, from 2001 until March 2024. He is currently a professor emeritus at the University of Osaka and a president and professor at the Graduate School of Management and Education Technology. His research interests include interactive computer graphics, human-computer interaction, and mixed reality.}
\end{biography}
\end{document}